\documentclass[
twocolumn,
]{ceurart}

\usepackage[table]{xcolor}
\newcommand{\good}[1]{\textbf{\textcolor{green!45!black}{#1}}}

\usepackage{listings}
\begin{document}

\copyrightyear{2026}
\copyrightclause{Copyright for this paper by its authors.
  Use permitted under Creative Commons License Attribution 4.0
  International (CC BY 4.0).}

\conference{RecSys in HR '26: The 6th Workshop on Recommender Systems for Human Resources, in conjunction with the 20th ACM Conference on Recommender Systems, September 28--October 2, 2026, Minneapolis, MN, United States.}

\title{Not All Matches Are Equally Valuable: An Online Experiment of Retention-Focused Recommendation in a Job-Matching Platform}

\author[1]{Tatsuya Ute}[email=tatsu@wantedly.com]
\address[1]{Wantedly, Inc., Tokyo, Japan.}

\author[1]{Chiaki Ichimura}[email=ichimura@wantedly.com]

\author[2]{Yuta Saito}[email=saito@hanjuku-kaso.com]
\address[2]{Hanjuku-kaso Co., Ltd., Tokyo, Japan.}

\begin{abstract}
Recommender systems in two-sided matching platforms are commonly optimized for immediate engagement signals such as click-through rate, reply rate, or the total number of successful matches. However, in real-world marketplaces, maximizing matches alone may be misaligned with business goals such as user churn rate and platform revenue, especially when users with fewer matches are at substantially higher risk of churn. In this paper, we study a real job matching platform and show that users with very few recent matches are indeed much more likely to leave the platform, while additional matches for already successful users provide limited marginal value for retention. Motivated by this empirical finding, we formulate a retention-aware recommendation problem and implement a simple post-processing method that adjusts the baseline match-focused ranking to prevent user churn. Specifically, the implemented method gives a score boost to churn-risk users with the goal of increasing their likelihood of obtaining matches and improving retention. We evaluate this practical approach in an online experiment on a real job-matching platform. The treatment group showed directionally lower user churn than the control group, although the estimated effect was not statistically significant at conventional levels, while company-side churn showed no evidence of deterioration. To our knowledge, this is among the first online experimental studies to investigate retention-focused recommendation in a real reciprocal job-matching platform.
\end{abstract}


\maketitle

\section{Introduction}
Recommender systems in two-sided matching platforms are often optimized for immediate engagement signals such as click-through rate, reply rate, or the total number of successful matches~\cite{Pizzato2010Reciprocal,Palomares2021Reciprocal,Yang2024RevisitingRRS}. This design is reasonable because these objectives are directly measurable and closely related to short-term platform activity. However, in real-world matching markets, optimizing only the total number of matches may not align with the platform's business objectives. This is because additional matches generated for users who have already achieved sufficient matching success often contribute little to retention, whereas users with few matches remain at high risk of churn and may benefit much more from additional matches \cite{Wu2017Returning,Wang2022Surrogate,kishimotobeyond}.

\begin{figure}
\centering
    \includegraphics[width=\linewidth]{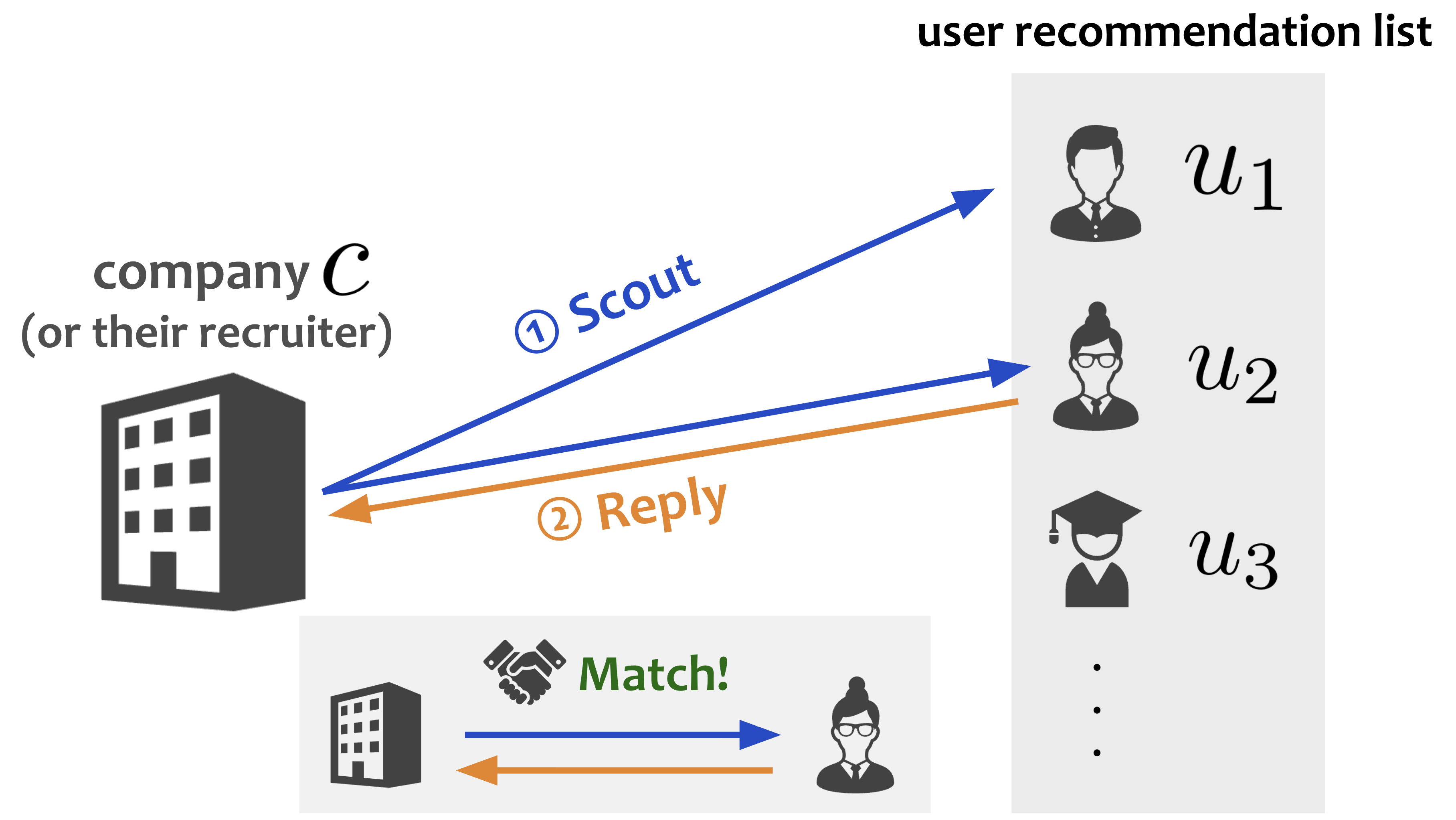}
    \caption{An example of the reciprocal recommendation setting considered in this study. The platform recommends a list of users for companies (or their recruiters), and companies then decide whether to send scouts to some of the users. Users who receive a scout then decide whether to respond. Users who receive a scout may in turn respond to it. Matches are recorded according to the platform's operational definition. We omit platform-specific details of the reply interface and match definition, as they are not essential to the setting studied here.}
    \label{fig:problem} 
   
\end{figure}

\begin{figure}
    \centering
    \includegraphics[width=\linewidth]{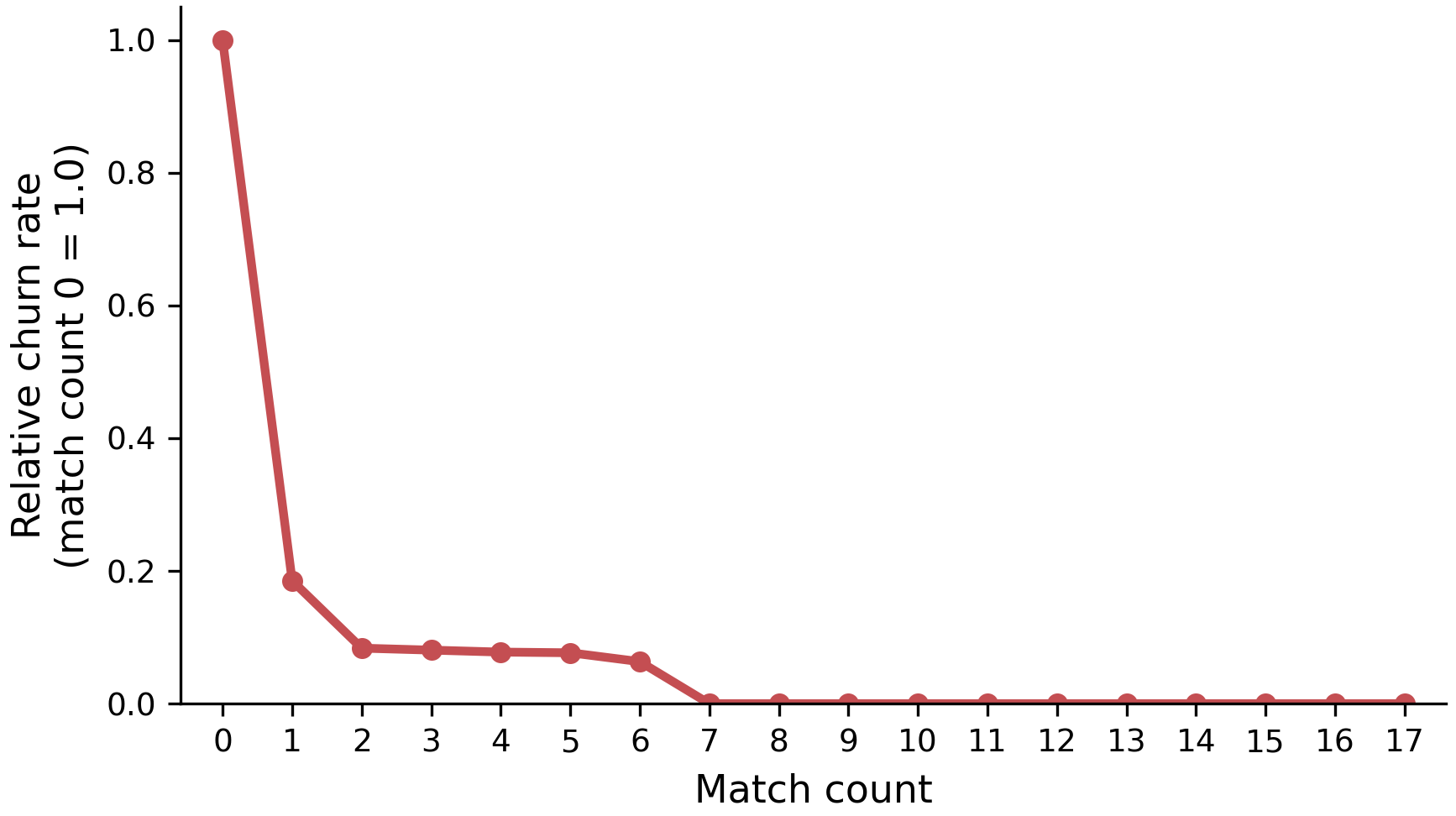}
    \caption{Relationship between users' recent match counts and churn risk. Users with very few matches are much more likely to churn, whereas the marginal retention gain from additional matches is limited for users who have already obtained several matches.}
    \label{fig:analysis}
\end{figure}

This paper is motivated by a case study of a job matching service, \textit{Wantedly Visit}\footnote{https://www.wantedly.com}, where a recruiter views recommended candidates in the context of a specific job opening and may send scouting messages to selected candidates (Figure~\ref{fig:problem} illustrates the platform's recommendation problem). A prior recommendation policy implemented on the platform ranked users for each company in descending order of predicted match probability estimated by a machine learning model. While this strategy was effective in maximizing the total number of matches, it also concentrated exposure on users who were already likely to match, leaving a substantial number of users with fewer matches underserved, a phenomenon that is broadly consistent with prior observations in marketplace recommendation and reciprocal recommendation \cite{Palomares2021Reciprocal,Acharya2023InactiveMembers,kishimotobeyond}. As shown in Figure~\ref{fig:analysis}, our analysis indeed revealed that users with very few recent matches were substantially more likely to churn, whereas providing additional matches to users who have already obtained sufficient matches had only limited marginal value for retention. This empirical observation suggests that the recommendation problem should not be framed solely as maximizing aggregate matches. Instead, the platform should explicitly account for \emph{retention risk} and prioritize users in a high-risk churn region. In other words, the key objective is not simply to create one more match anywhere in the platform, but to allocate opportunities so as to reduce the number of users whose match count remains below a critical churn-risk threshold.

Based on this insight, we implement a simple retention-focused post-processing strategy on top of the baseline match-probability ranking. Specifically, for relevant users judged to be at risk of churn, the method gives a score boost, thereby allocating them greater exposure than they would otherwise receive under the baseline ranking. For other users, the baseline ranking is left unchanged. This lightweight intervention aims to improve user retention without sacrificing other key business metrics.

We conducted a large-scale online experiment on \textit{Wantedly Visit}, a real job-matching platform. The treatment group exhibited directionally lower user churn than the match-maximizing baseline, although the estimated difference was statistically inconclusive at conventional levels, while we found no evidence of deterioration in company-side churn. To the best of our knowledge, this study provides one of the first online experimental investigations of retention-focused recommendation in a real reciprocal job-matching platform.

\section{Related Work}
Our work is related to several lines of research in recommender systems and marketplace design.
First, our study is related to the literature on recommender systems for matching markets and reciprocal recommendation. In such environments, a successful interaction depends on the preferences and behaviors of both sides, and naive one-sided relevance optimization can produce inefficient or unbalanced outcomes. Prior work has shown that explicitly modeling both sides is essential for improving matching quality and marketplace health \cite{Palomares2021Reciprocal, Pizzato2010Reciprocal, Yang2024RevisitingRRS}.

Second, our work is connected to long-term and retention-aware recommendation. While many industrial recommender systems optimize short-term engagement, prior studies have emphasized the importance of delayed outcomes such as satisfaction, continued participation, and long-term user value \cite{Wu2017Returning, Wang2022Surrogate, Acharya2023MemberValue, Acharya2023InactiveMembers, kishimotobeyond, Jannach2022MultiObjective}. Our approach belongs to this family of methods but differs in two respects. First, it emphasizes a simple operational intervention that prioritizes users whose recent match volume places them in a high-risk churn region, and second, it evaluates this intervention through an online experiment on a real two-sided matching platform.

Finally, our method is also related to exposure allocation and fairness in recommendation \cite{Singh2018FairnessExposure, Vassoy2024ConsumerSideFairness, Tomita2024FairReciprocal, Zhao2023FairnessDiversity, Zehlike2022FairnessSurvey, Singh2019PolicyLearningFairness, Morik2020ControllingFairnessBias}. Methods such as FairRec~\cite{patro2020fairrec} and FEIR~\cite{li2024feir} explicitly counteract concentration induced by relevance-only ranking by considering the distribution of exposure, utility, or competitive disadvantage across both sides of a marketplace. Our objective differs from these fairness-oriented approaches, i.e., rather than optimizing an exposure-parity constraint or another formal fairness criterion, we allocate a limited amount of additional exposure according to an empirically defined retention-risk signal.

\section{Problem Formulation}
We consider a two-sided job matching platform with a set of users $\mathcal{U}$ and a set of companies $\mathcal{C}$. For each user $u \in \mathcal{U}$ and company $c \in \mathcal{C}$\footnote{Depending on the platform implementation, $c$ may represent either a recruiter evaluating candidates or a particular job opening. Whether $c$ is defined at the recruiter or job-opening level does not affect our formulation, main contributions, or conclusions.}, the system estimates a match probability
\[
p_{\mathrm{match}}(u,c) = p_{\mathrm{scout}}(u,c) \cdot p_{\mathrm{reply}}(u,c) \in [0,1].
\]
Here, $p_{\mathrm{scout}}(u,c)$ denotes the probability that company $c$ sends a scout to user $u$ after the user is recommended, and $p_{\mathrm{reply}}(u,c)$ denotes the conditional probability that user $u$ replies to the scout from company $c$, given that a scout is sent.  Their product $p_{match}(u,c)$ therefore represents the probability that recommending user $u$ to company $c$ leads to a successful match as operationally defined above.

In the standard formulation, including our previous recommender, the system ranks users for each company in descending order of $p_{\mathrm{match}}(u,c)$ and optimizes the expected total number of matches,
\begin{align}
    \max_{\pi} \sum_{c \in \mathcal{C}} \sum_{u \in \pi(c)} p_{\mathrm{match}}(u,c), \label{eq:max_match}
\end{align}
where $\pi(c)$ denotes the ordered list of recommended users for company (or recruiter) $c$.

Although this objective is reasonable from the perspective of short-term engagement, it does not capture user retention. Let $m_u(\pi)$ denote the number of matches obtained by user $u$ under $\pi$ in a predefined observation window. The empirical analysis for the target service, shown in Figure~\ref{fig:analysis}, suggested that users with very small values of $m_u(\pi)$ were substantially more likely to churn, whereas the marginal retention gain from increasing $m_u(\pi)$ for already successful users was limited. A similar trend was recently reported using real-world data from an online dating platform by \citet{kishimotobeyond}.\footnote{Unlike our work, \citet{kishimotobeyond} did not evaluate their retention-focused recommendation method through an online experiment on a real-world matching platform.} Thus, in a real matching platform, there often exists a churn-risk threshold $\tau$, defined as the minimum recent match count above which churn risk decreases substantially, such that users with $m_u(\pi) < \tau$ are considered to be at risk of churn.

This observation motivates a different optimization perspective compared to Eq.~\eqref{eq:max_match}. Instead of treating every additional expected match equally, we seek to reduce the number of users whose match volume remains below the critical threshold $\tau$. More specifically, let
\[
\mathrm{churn\_risk}(u;\pi) = \mathbb{I}[m_u(\pi) < \tau]
\]
be an indicator of whether user $u$ is at risk of churn.
Then, the platform-level objective can be formulated as minimizing the number of churn-risk users as follows.
\begin{align}
    \min_{\pi} \sum_{u \in \mathcal{U}} \mathrm{churn\_risk}(u;\pi)
\end{align}
The challenge is that the recommendation policy must strategically prioritize users who are at risk of churn to optimize this objective, even when they are not the users with the highest immediate match probability for companies. We examine this interesting trade-off between user-side churn and company-side satisfaction in the experiment section (Section 5).

\begin{figure}[t]
\centering
\includegraphics[width=\linewidth]{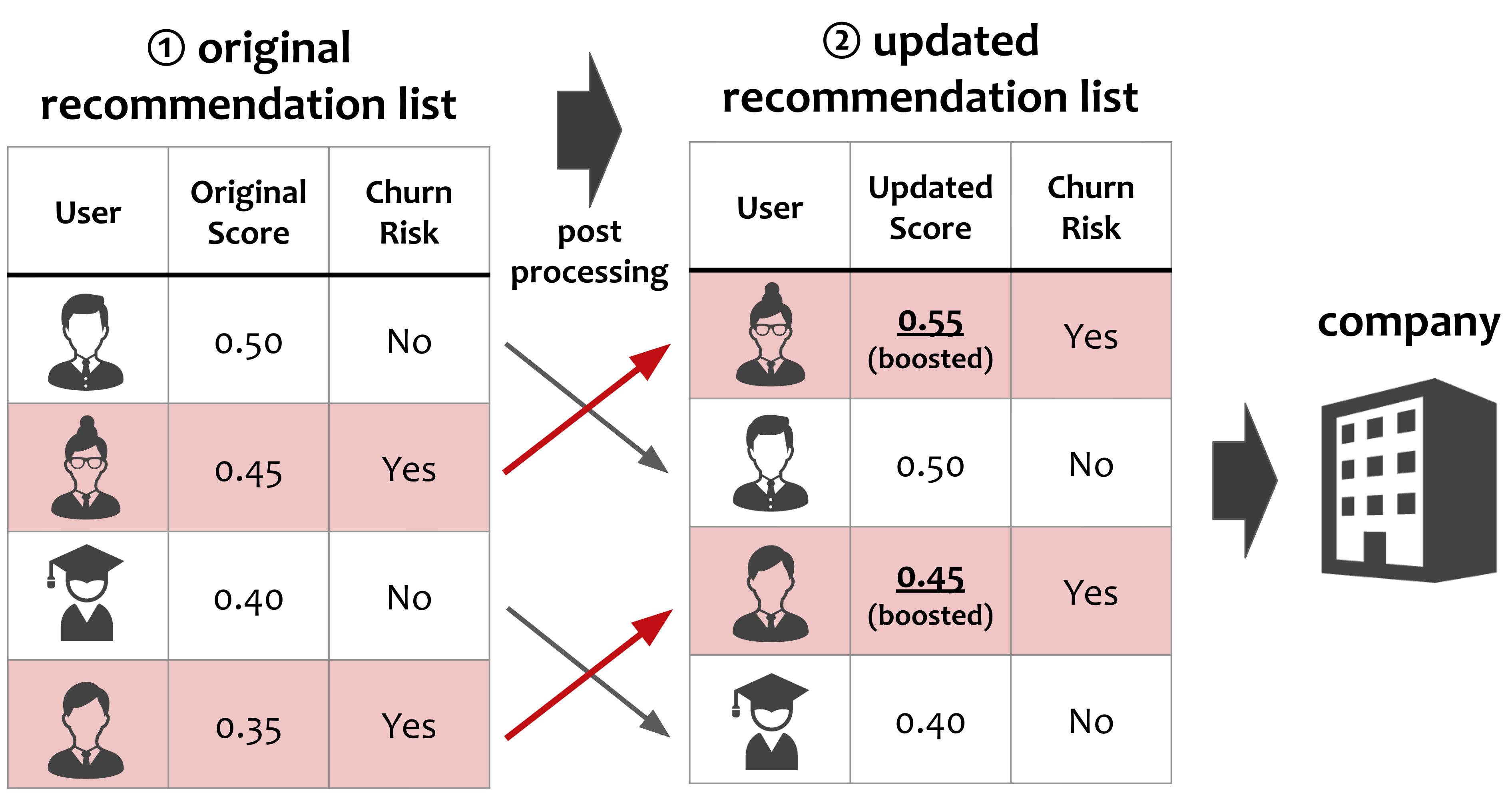}
\caption{Illustration of the proposed retention-focused post-processing strategy. Starting from the original recommendation list produced by a baseline model, the method boosts the scores of users identified as being at risk of churn, while leaving other users unchanged. As a result, some churn-risk users move upward in the recommendation list after post-processing, increasing their chance of being exposed and obtaining a match.}
\label{fig:retention_strategy}
\vspace{-2mm}
\end{figure}

\section{Implemented Retention-Focused Strategy} \label{sec:proposed}
This section describes our practical post-processing intervention for retention-focused recommendations. Rather than replacing the baseline recommender with an entirely new optimization procedure, we implement retention-focused recommendation as a lightweight post-processing step on top of the existing match-probability ranking model. This design choice keeps the method operationally simple and allows smooth deployment on the production system.

Let $s(u,c)$ denote the baseline recommendation score for user $u$ and company $c$, which is often derived from the predicted match probability. Our method modifies this score only for users who are judged to be at risk of churn. Section 3 formulates retention risk in terms of the recent match count $m_u$. Since a match requires both a scout and a reply, the stage that the ranking most directly affects is scout receipt: it determines which candidates a recruiter sees. We therefore use $s_u$, the number of scouts a user received in the recent window, as the operational risk signal in the production system. A churn prediction model fit on historical log data with a range of pre-treatment behavioral features identified low recent scout receipt as a leading indicator of churn, and we accordingly treated users with fewer than three recent scouts as the high-risk region, $\mathcal{U}_{\mathrm{risk}} = \{u \in \mathcal{U} \mid s_u < 3\}$. This cutoff was chosen from that analysis rather than through a formal breakpoint test, and the underlying relationship is predictive rather than causal; it should be interpreted as a platform-specific design choice rather than a universal churn threshold.

For users in $\mathcal{U}_{\mathrm{risk}}$, we apply a score boost to give more exposure to those users. More specifically, let $\mathrm{score\_cutoff}$ denote the minimum baseline score required to apply the boost, and let $n(u)$ denote the number of times user $u$ has appeared in top-ranked positions (up to a predefined ranking cutoff) across rankings for all companies. Then, for each $(u,c)$-pair, we update the score as
\[
s'(u,c)=
\left\{\;\begin{aligned}
&\mathrm{clip}\bigl(s(u,c)+\beta\gamma^{\,n(u)},-1,1\bigr),\\[-0.5mm]&\quad \text{if }u \in \mathcal{U}_{\mathrm{risk}}\text{ and }s(u,c)>\mathrm{score\_cutoff},\\[2mm]
&s(u,c),\qquad \text{otherwise}.
\end{aligned}\right.
\]
where $\beta$ is the base boost factor, $\gamma$ is the diversity decay factor, and $\mathrm{clip}(\cdot,-1,1)$ truncates the final score to the valid range. The values of $\beta$ and $\gamma$ can be selected through offline parameter tuning to optimize a predicted retention objective. After the post-processing step, users are ranked in descending order of the modified score $s'(u,c)$. Figure~\ref{fig:retention_strategy} illustrates the intuition of the proposed post-processing intervention.

This post-processing rule has three important properties. First, the boost is applied only to churn-risk users, reflecting our retention-oriented objective. Second, churn-risk users are not boosted indiscriminately; the boost is applied only when the predicted match probability for a $(u,c)$-pair is at least $\mathrm{score\_cutoff}$ because boosting is of little value if the pair is unlikely to match. Third, the boost decays with the user's overall exposure across companies, which prevents the method from excessively concentrating the boost on users who are already widely visible in the platform's rankings. Thus, the method increases exposure for users who are at risk of churn but still have a reasonable likelihood of obtaining matches, while avoiding severe congestion in the recommendation space.
 
The main advantage of this approach is its practicality. Because the implemented method only adds a small adjustment to sufficiently promising users and leaves the rest of the ranking unchanged, it does not require retraining the underlying recommendation model or directly optimizing a complex retention objective. Instead, it introduces an interpretable and readily deployable intervention that strategically biases the ranking toward users with greater retention need.

\section{Online Experiment}
We conducted a large-scale online experiment on a recommender system implemented on \textit{Wantedly Visit} to evaluate whether retention-focused recommendation can reduce user churn without sacrificing other key business indicators.

\subsection{Experiment Design}
The online experiment was conducted on the platform between November 10, 2025, and December 8, 2025. We randomized at the recruiter level (with each company potentially having multiple recruiters), resulting in approximately 3,000 recruiters per group. The analysis focused on users who had been active\footnote{A user is considered active on a given day if at least one login to the platform is recorded on that day.} on at least one day during the 28-day pre-period (2025/10/13--2025/11/09) and received at least one impression from either a control or a treatment recruiter during the test period. Under this definition, the population consisted of approximately 20.9k user observations in the control-exposed group and 20.4k in the treatment-exposed group. Because randomization was performed at the recruiter level while the primary outcome was measured at the user level, a user could in principle be exposed to both groups. We therefore report a primary analysis allowing such overlap, as well as a robustness analysis restricted to users who were exposed to only one experiment group; in the latter sample, the analysis included approximately 6.6k control-only users and 6.0k treatment-only users.\footnote{Note that the reported user and recruiter counts should be interpreted as approximate sample sizes for the AB-test analysis population, not as platform-wide totals.}

The baseline policy followed the conventional match-maximization strategy described in Eq.~\eqref{eq:max_match}, whereas the treatment policy applied the post-processing method presented in Section~\ref{sec:proposed}, which gives a score boost for churn-risk users while applying a smaller boost to users who already appear frequently in recruiters' rankings. Our primary outcome was a binary churn flag, that is, among users who were active at least once during the pre-period, a user was considered to have churned if they did not become active at least once during the test period. 

To estimate the treatment effect on user churn, we fit a user-level logistic regression with the treatment assignment indicator and 14 pre-treatment covariates. All continuous covariates were standardized before fitting the model. The covariates covered user attributes and preference settings, such as student status, scout acceptance preferences, and willingness for career change or side jobs; recent behavioral signals, such as the number of applications, matches, and received scouts; exposure volume, such as the number of ranking impressions; and activity-history variables, such as the number of days since the last active date. Because treatment was randomized, the coefficient on the treatment indicator can be interpreted as the average treatment effect up to the potential contamination induced by cross-group exposure, while the covariates improve estimation precision. 

In addition to the churn outcome, we examined company-side guardrail metrics, including actual company churn and a model-based company churn score, to verify that improving user retention did not come at the expense of company-side outcomes. To the best of our knowledge, this is among the first online experiments that directly compare a retention-focused recommendation policy with a match-maximization policy in a real two-sided matching platform.

\begin{table}[t]
\centering
\caption{Effects of the implemented post-processing method on user churn. Bold values indicate estimates in the intended direction. None of the estimates is statistically significant at conventional levels.}
\label{tab:regression}
\begin{tabular}{lcc}
\toprule
Sample / subgroup & Effect (Odds Ratio) & $p$-value \\
\midrule
\multicolumn{3}{l}{\textbf{Panel A: Main analysis}} \\
Overlap allowed &  \good{$0.957$} & 0.128 \\
Mutually exclusive only   & \good{$0.948$} & 0.223 \\
\midrule
\multicolumn{3}{l}{\textbf{Panel B: By prior number of received scouts}} \\
0 scouts  & \good{$0.956$} & 0.200 \\
1--2 scouts   & \good{$0.948$} & 0.359 \\
3+ scouts & $1.015$ & 0.893 \\
\midrule
\multicolumn{3}{l}{\textbf{Panel C: By prior number of impressions}} \\
0 impressions  & \good{$0.845$} & 0.139 \\
1--5 impressions   & \good{$0.903$} & 0.095 \\
6--20 impressions & $0.973$ & 0.564 \\
21+ impressions & $1.013$ & 0.817 \\
\bottomrule
\end{tabular}
\vspace{-2mm}
\end{table}

\subsection{Results}
This section reports and discusses the results of the online experiment.

Panel A of Table~\ref{tab:regression} summarizes the regression results for the user churn outcome. In both specifications, the estimated effect of the implemented post-processing method is in the desired direction: the odds ratio is below 1, indicating lower estimated odds of churn in the treatment group than in the control group. In the primary specification allowing overlap, the estimated odds ratio is 0.957, and in the mutually exclusive robustness sample, the odds ratio is 0.948. Although neither estimate is statistically significant at conventional levels, the point estimates are directionally consistent across specifications. We therefore interpret these results as directionally positive but statistically inconclusive evidence regarding the effect of the intervention on user churn.\footnote{Our product decision criterion does not rely exclusively on the conventional p<0.05 threshold. Recent work on experimentation programs argues that product decisions may also account for estimated effect sizes, opportunity costs, and other business considerations~\cite{Sudijono2024OptimizingReturns,Chou2025EvaluatingDecisionRules}. Importantly, this decision criterion is distinct from statistical inference: the present estimates should be regarded as statistically inconclusive under conventional significance thresholds. We therefore separate our practical deployment decision from the strength of the statistical evidence reported in this paper.}

We next conducted exploratory subgroup analyses to examine whether the estimated effects were directionally consistent with the intended target population (Panel B in Table~\ref{tab:regression}). When users are grouped by the number of scouts received during the pre-period (Panel B in Table 1), the point estimates are more favorable among users with 0–2 prior scouts: the odds ratio is 0.956 for users with 0 prior scouts and 0.948 for users with 1–2 prior scouts, compared with 1.015 for users with 3 or more prior scouts. Panel C shows a similar descriptive pattern by prior impressions: the odds ratio is 0.845 for users with 0 prior impressions and 0.903 for those with 1–5 prior impressions, compared with 0.973 for users with 6–20 impressions and 1.013 for those with 21 or more impressions. These exploratory patterns are consistent with the intended targeting of underexposed users, but none of the subgroup estimates is statistically significant at conventional levels, and they should not be interpreted as evidence of treatment-effect heterogeneity.

We next examined overall matching outcomes. After covariate adjustment using Poisson regression with company-level controls, the point estimate for the total number of matches was 5.4\% higher in the treatment group than in the control group. However, this difference was not statistically significant, and we therefore do not interpret it as evidence that the intervention increased overall matching performance. Importantly, the retention-focused intervention was not designed to increase the total number of matches, and the observed positive point estimate should be regarded as exploratory. One possible interpretation is that redistributing recruiter attention may alleviate congestion caused by repeated exposure of the same users and thereby affect conversion opportunities, consistent with prior work on congestion in job recommendation settings~\cite{Mashayekhi2023ReCon,Su2022OptimizingRankings}. 

We also assessed whether the proposed method harmed company-side retention. For this purpose, we analyzed companies whose contracts ended during the observation window among companies participating in the online experiment. Using weighted least squares with company-clustered robust standard errors, we found no evidence of deterioration in either actual company churn or a model-based company churn score: the estimated treatment–control coefficient difference was \(+0.006\) for actual churn (\(p=0.77\)) and \(-0.007\) for the churn score (\(p=0.34\)). Hence, within the precision of the current experiment, the user-side gains were not achieved at the expense of company-side outcomes.

\section{Limitations}
This study has several limitations. First, the experiment was conducted on a single job-matching platform over a four-week period, and the results may not generalize to other marketplaces, recommendation interfaces, or longer-term retention outcomes.

Second, randomization occurred at the recruiter level, while outcomes were measured at the user level, and some users received exposure from recruiters assigned to both experimental arms. Although we report a mutually exclusive-user analysis as a robustness check, cross-arm exposure limits a clean user-level causal interpretation.

Third, the estimated effects on the primary churn outcome were not statistically distinguishable from the null with respect to $p<0.05$, and the subgroup analyses were exploratory and insufficiently powered to establish treatment-effect heterogeneity. We also did not directly verify the mechanism through which the reranking intervention may affect retention. In particular, we did not separately establish whether boosted users received more effective exposure, scouts, or matches, nor whether changes in these intermediate outcomes mediated the estimated churn effect. Therefore, the observed churn estimates and the increase in aggregate matches should not be interpreted as causal evidence for a specific mediation pathway.

Fourth, the churn-risk threshold and score-boosting parameters were chosen from offline analyses rather than learned through a fully specified retention objective. Figure 2 is observational and does not strictly establish a causal effect of additional matches on churn.

Fifth, we compared the intervention only with the existing production baseline. We did not perform a head-to-head comparison with alternative congestion-aware or fairness-aware allocation methods that are relevant to our work; consequently, the experiment does not establish superiority over those approaches. Rather, the experiment is intended to illustrate the practical potential of the implemented retention-focused approach for a real matching platform.

Our intervention also raises fairness and transparency considerations. Because the method intentionally reallocates exposure toward users identified as being at risk of churn, some users with higher baseline match scores may receive less exposure than they would under a match-maximizing ranking, creating a potential trade-off among user retention, candidate-side exposure, and company-side utility. Although our company-side guardrail analysis found no evidence of deterioration in company churn, this does not by itself establish that the intervention is fair or that recruiters and users would perceive the policy as acceptable. We did not directly evaluate perceived fairness, transparency, or reactions to disclosure of the reranking policy; examining these aspects, as well as possible fairness-aware constraints, is an important direction for future work. Finally, privacy and commercial constraints limit the release of production logs and parts of the baseline model.

\section{Conclusion and Future Directions}
In this paper, we seek to design practical recommender systems that improve user retention rather than maximize total matches or satisfy fairness objectives. In the target platform, maximizing predicted matches alone was not well aligned with the business goal, because users with too few matches were more likely to churn. To address this issue, we formulated a retention-aware ranking problem and implemented a simple post-processing method that adjusts the baseline match-focused ranking for churn-risk users through a small, controlled score boost. The proposed post-processing method yielded directionally lower user churn in the online experiment, although the estimated effect was statistically inconclusive at conventional levels. These results provide an initial industrial case study of retention-focused recommendation in a real reciprocal job-matching platform. More broadly, this case study suggests that marketplace recommenders should be carefully optimized for the true downstream objective rather than only for an immediate proxy like the number of matches.

As future work, we plan to extend the framework to optimize company retention as well as user retention, since company-side retention is more directly linked to platform revenue through mechanisms such as fees for job postings and user messaging. One possible approach is to apply a similar score-boosting intervention to systems that recommend companies or job postings to users. We also plan to improve the identification of users at risk of churn by incorporating a broader range of signals, as churn risk may depend on factors beyond the recent number of matches.

\section*{Declaration on Generative AI}
During the preparation of this work, the authors used Claude Opus 4.8, Claude Opus 5, GPT-5.5 and GPT-5.6 Sol in order to: Grammar and spelling check, Paraphrase and reword, Drafting content. After using these tools, the authors reviewed and edited the content as needed and take full responsibility for the publication's content.

\bibliography{ref}

@article{Jannach2022MultiObjective,
  author       = {Dietmar Jannach},
  title        = {Multi-Objective Recommender Systems: Survey and Challenges},
  journal      = {CoRR},
  volume       = {abs/2210.10309},
  year         = {2022},
  url          = {https://arxiv.org/abs/2210.10309}
}

@article{Zhao2023FairnessDiversity,
  author       = {Yuying Zhao and Yu Wang and Yunchao Liu and Xueqi Cheng and Charu C. Aggarwal and Tyler Derr},
  title        = {Fairness and Diversity in Recommender Systems: A Survey},
  journal      = {ACM Transactions on Intelligent Systems and Technology},
  year         = {2024},
  note         = {to appear / survey version available on arXiv:2307.04644},
  url          = {https://arxiv.org/abs/2307.04644}
}

@article{Palomares2021Reciprocal,
  author       = {Iv{\'a}n Palomares and Carlos Porcel and Luiz Pizzato and Ido Guy and Enrique Herrera-Viedma},
  title        = {Reciprocal Recommender Systems: Analysis of State-of-Art Literature, Challenges and Opportunities Towards Social Recommendation},
  journal      = {Information Fusion},
  volume       = {69},
  pages        = {103--127},
  year         = {2021},
  doi          = {10.1016/j.inffus.2020.12.001}
}

@inproceedings{Pizzato2010Reciprocal,
  author       = {Luiz Pizzato and Tomek Rej and Thomas Chung and Irena Koprinska and Judy Kay},
  title        = {Reciprocal Recommender System for Online Dating},
  booktitle    = {Proceedings of the 4th ACM Conference on Recommender Systems},
  series       = {RecSys '10},
  pages        = {207--214},
  year         = {2010},
  doi          = {10.1145/1864708.1864787}
}

@inproceedings{Yang2024RevisitingRRS,
  author       = {Chen Yang and Sunhao Dai and Yupeng Hou and Wayne Xin Zhao and Jun Xu and Yang Song and Hengshu Zhu},
  title        = {Revisiting Reciprocal Recommender Systems: Metrics, Formulation, and Method},
  booktitle    = {Proceedings of the 30th ACM SIGKDD Conference on Knowledge Discovery and Data Mining},
  series       = {KDD '24},
  pages        = {3714--3723},
  year         = {2024},
  doi          = {10.1145/3637528.3671734}
}

@inproceedings{Wu2017Returning,
  author       = {Qingyun Wu and Hongning Wang and Liangjie Hong and Yue Shi},
  title        = {Returning is Believing: Optimizing Long-Term User Engagement in Recommender Systems},
  booktitle    = {Proceedings of the 26th ACM International Conference on Information and Knowledge Management},
  series       = {CIKM '17},
  pages        = {1927--1936},
  year         = {2017},
  doi          = {10.1145/3132847.3133025}
}

@inproceedings{Wang2022Surrogate,
  author       = {Yuyan Wang and Mohit Sharma and Can Xu and Sriraj Badam and Qian Sun and Lee Richardson and Lisa Chung and Ed H. Chi and Minmin Chen},
  title        = {Surrogate for Long-Term User Experience in Recommender Systems},
  booktitle    = {Proceedings of the 28th ACM SIGKDD Conference on Knowledge Discovery and Data Mining},
  series       = {KDD '22},
  pages        = {4100--4109},
  year         = {2022},
  doi          = {10.1145/3534678.3539073}
}

@inproceedings{Singh2018FairnessExposure,
  author       = {Ashudeep Singh and Thorsten Joachims},
  title        = {Fairness of Exposure in Rankings},
  booktitle    = {Proceedings of the 24th ACM SIGKDD International Conference on Knowledge Discovery and Data Mining},
  series       = {KDD '18},
  pages        = {2219--2228},
  year         = {2018},
  doi          = {10.1145/3219819.3220088}
}

@article{Vassoy2024ConsumerSideFairness,
  author       = {Bo Vass{\o}y and Asi Shekari and Katrien Verbert},
  title        = {Consumer-side Fairness in Recommender Systems: A Survey},
  journal      = {Artificial Intelligence Review},
  year         = {2024},
  doi          = {10.1007/s10462-023-10663-5}
}

@inproceedings{Tomita2024FairReciprocal,
  author       = {Yoji Tomita and Tomohiko Yokoyama},
  title        = {Fair Reciprocal Recommendation in Matching Markets},
  booktitle    = {Proceedings of the 18th ACM Conference on Recommender Systems},
  series       = {RecSys '24},
  year         = {2024},
  doi          = {10.1145/3640457.3688130}
}

@inproceedings{kishimotobeyond,
  title={Beyond Match Maximization and Fairness: Retention-Optimized Two-Sided Matching},
  author={Kishimoto, Ren and Takehi, Rikiya and Tanaka, Koichi and Tomita, Yoji and Nomura, Masahiro and Togashi, Riku and Saito, Yuta},
  booktitle={International Conference on Learning Representations},
  volume={2026},
  pages={80483--80501},
  year={2026}
}

@inproceedings{Acharya2023MemberValue,
  author    = {Ayan Acharya and Siyuan Gao and Ankan Saha and Borja Ocejo and Kinjal Basu and Keerthi Selvaraj and Rahul Mazumdar},
  title     = {Optimizing for Member Value in an Edge Building Marketplace},
  booktitle = {Proceedings of the 32nd ACM International Conference on Information and Knowledge Management},
  series    = {CIKM '23},
  pages     = {5--14},
  year      = {2023},
  doi       = {10.1145/3583780.3615000}
}

@inproceedings{Acharya2023InactiveMembers,
  author    = {Ayan Acharya and Siyuan Gao and Borja Ocejo and Kinjal Basu and Ankan Saha and Keerthi Selvaraj and Rahul Mazumdar and Parag Agrawal and Aman Gupta},
  title     = {Promoting Inactive Members in Edge-Building Marketplace},
  booktitle = {Companion Proceedings of the ACM Web Conference 2023},
  series    = {WWW '23 Companion},
  pages     = {945--949},
  year      = {2023},
  doi       = {10.1145/3543873.3587647}
}

@inproceedings{Su2022OptimizingRankings,
  author    = {Yi Su and Magd Bayoumi and Thorsten Joachims},
  title     = {Optimizing Rankings for Recommendation in Matching Markets},
  booktitle = {Proceedings of the ACM Web Conference 2022},
  series    = {WWW '22},
  year      = {2022},
  pages     = {328--338},
  publisher = {Association for Computing Machinery},
  address   = {New York, NY, USA},
  doi       = {10.1145/3485447.3511961}
}

@inproceedings{Mashayekhi2023ReCon,
  author    = {Yoosof Mashayekhi and Bo Kang and Jefrey Lijffijt and Tijl De Bie},
  title     = {ReCon: Reducing Congestion in Job Recommendation using Optimal Transport},
  booktitle = {Proceedings of the 17th ACM Conference on Recommender Systems},
  series    = {RecSys '23},
  year      = {2023},
  pages     = {696--701},
  publisher = {Association for Computing Machinery},
  address   = {New York, NY, USA},
  doi       = {10.1145/3604915.3608817}
}

@article{Zehlike2022FairnessSurvey,
  author  = {Meike Zehlike and Ke Yang and Julia Stoyanovich},
  title   = {Fairness in Ranking: A Survey},
  journal = {ACM Computing Surveys},
  year    = {2022},
  doi     = {10.1145/3533379}
}

@article{Singh2019PolicyLearningFairness,
  author  = {Ashudeep Singh and Thorsten Joachims},
  title   = {Policy Learning for Fairness in Ranking},
  journal = {Advances in Neural Information Processing Systems},
  volume  = {32},
  year    = {2019}
}

@inproceedings{Morik2020ControllingFairnessBias,
  author    = {Marco Morik and Ashudeep Singh and Jessica Hong and Thorsten Joachims},
  title     = {Controlling Fairness and Bias in Dynamic Learning-to-Rank},
  booktitle = {Proceedings of the 43rd International ACM SIGIR Conference on Research and Development in Information Retrieval},
  series    = {SIGIR '20},
  pages     = {429--438},
  year      = {2020},
  publisher = {Association for Computing Machinery},
  doi       = {10.1145/3397271.3401100}
}

@article{Sudijono2024OptimizingReturns,
  author  = {Timothy Sudijono and Simon Ejdemyr and Apoorva Lal and Martin Tingley},
  title   = {Optimizing Returns from Experimentation Programs},
  journal = {arXiv preprint arXiv:2412.05508},
  year    = {2024},
  doi     = {10.48550/arXiv.2412.05508},
  url     = {https://arxiv.org/abs/2412.05508}
}

@inproceedings{Chou2025EvaluatingDecisionRules,
  author    = {Winston Chou and Colin Gray and Nathan Kallus and Aur{\'e}lien Bibaut and Simon Ejdemyr},
  title     = {Evaluating Decision Rules Across Many Weak Experiments},
  booktitle = {Proceedings of the 31st ACM SIGKDD Conference on Knowledge Discovery and Data Mining V.2},
  series    = {KDD '25},
  year      = {2025},
  pages     = {4365--4374},
  publisher = {Association for Computing Machinery},
  doi       = {10.1145/3711896.3737217}
}

@inproceedings{patro2020fairrec,
  author    = {Gourab K. Patro and
               Arpita Biswas and
               Niloy Ganguly and
               Krishna P. Gummadi and
               Abhijnan Chakraborty},
  title     = {{FairRec}: Two-Sided Fairness for Personalized Recommendations
               in Two-Sided Platforms},
  booktitle = {Proceedings of The Web Conference 2020},
  pages     = {1194--1204},
  year      = {2020},
  publisher = {Association for Computing Machinery},
  doi       = {10.1145/3366423.3380196}
}

@article{li2024feir,
  author  = {Nan Li and
             Bo Kang and
             Jefrey Lijffijt and
             Tijl De Bie},
  title   = {{FEIR}: Quantifying and Reducing Envy and Inferiority for Fair
             Recommendation of Limited Resources},
  journal = {ACM Transactions on Intelligent Systems and Technology},
  volume  = {15},
  number  = {4},
  pages   = {80:1--80:24},
  year    = {2024},
  doi     = {10.1145/3643891}
}

\end{document}